# Development of an HPLC Analytical Method Based on Large Language Models: A Case Study of the Eczema External Wash Formula

**Kexin Guo[1+], Xueting Sun[1+], Xinyan Zhao[2,4], Wei Wang[1,3]， Xingchu Gong[1,2,4,5*]**

1. School of pharmacy, Zhejiang University, Hangzhou 310058, China

2. Institute of Traditional Chinese Medicine, Tianjin University of Traditional Chinese Medicine, Tianjin 301617, China

3. Department of Pharmacy, Hangzhou Third People' s Hospital, Hangzhou, Zhejiang, China

4.Jinhua Institute of Zhejiang University, Jinhua 321016, China

5. National Key Laboratory for the Creation of Modern Traditional Chinese Medicine, Hangzhou 310058, China

* Corresponding author: gongxingchu@zju.edu.cn; Tel.: +86-571-88208427

+ These authors contributed equally to this work

**Abstract**

Traditional Chinese medicine (TCM) formulas feature complex chemical compositions, and their HPLC quality control method development is time-consuming and experience-dependent. Large language models (LLMs) provide a promising strategy for intelligently assisting analytical method construction. Taking the Eczema External Wash Formula as the research object, this work constructed a human–machine collaborative workflow involving intelligent recommendation, experimental validation and iterative optimization for HPLC method development. Among five prompt strategies tested, expert-role setting with structured reasoning performed best. After inputting formula information, the LLM proposed five chromatographic schemes. Through experimental screening and optimization, the final HPLC method used a SunShell C18 column, mobile phase of 0.1% aqueous formic acid–acetonitrile for gradient elution, with seven markers separated within 55 min. Twenty components were characterized by LC-MS. The validated method exhibited satisfactory linearity, precision, repeatability, stability and recovery. This indicates LLMs can reduce the workload of chromatographic condition exploration and boost the development efficiency of TCM formula analytical methods, supporting intelligent TCM quality control.

## 1 Introduction

High-performance liquid chromatography (HPLC) is the technique of choice for detecting chemical constituents in traditional Chinese medicines (TCMs). However, HPLC method development still relies on trial-and-error approaches[1], which are inefficient and demand extensive personal experience. It is therefore necessary to find a more efficient route for developing HPLC methods for TCMs.

Large language models (LLMs), one of the most breakthrough technologies in the field of artificial intelligence in recent years, have greatly accelerated chemical synthesis[2–4], drug design, protein function prediction[5], and material property optimization[6]. Their ability to process textual, unstructured knowledge embedded in large volumes of literature, pharmacopoeias, and experimental reports makes them promising intelligent bridges linking the complex theoretical knowledge of TCM with modern chromatographic practice. In recent years, researchers have begun to explore the potential of LLMs in chromatographic retention-time prediction and condition recommendation[3,7,8]; by integrating the physicochemical properties of compounds with domain knowledge, they have demonstrated the feasibility of assisting experimental design[9], offering a new approach for the rapid development of TCM analytical methods.

The Eczema External Wash Formula is composed of seven herbs: Saposhnikoviae Radix, Phellodendri Cortex, Kochiae Fructus, Dictamni Cortex, Schizonepetae Herba, Sophorae Flavescentis Radix, and Cnidii Fructus. It is used clinically for acute and subacute eczema, contact dermatitis, and neurodermatitis. No analytical method for this formula has been reported in the literature. By comparing different prompt strategies and structuring the input of formula information and experimental conditions, this study generated candidate schemes with an LLM and optimized them iteratively through experiments, finally identified the chemical constituents by LC-MS, and established a multi-component quantitative method. The aim was to construct a human–machine collaborative development model of LLM-assisted prediction, experimental verification, and iterative optimization, providing a reference for the rapid development of analytical methods for TCM formulas.

## 2 Materials and Instruments

Reagents: berberine hydrochloride (batch No. JB241108; purity ≥98%; Shanghai Yuanye Bio-Technology Co., Ltd.); dictamnine (batch No. MUST-26042405; purity ≥99.9%; Shanghai Yaji

Biotechnology Co., Ltd.); momordin Ic (batch No. JB240100; purity ≥98%; Shanghai Yuanye Bio-Technology Co., Ltd.); osthole (batch No. MUST-25102507; purity ≥99.78%; Shanghai Yaji Biotechnology Co., Ltd.); prim-O-glucosylcimifugin (batch No. C18972035; purity ≥98%; Shanghai Macklin Biochemical Co., Ltd.); magnoflorine (batch No. D2511620; purity ≥98%; Shanghai Aladdin Biochemical Technology Co., Ltd.); imperatorin (batch No. LS444327; purity ≥98%; Shanghai Yuanye Bio-Technology Co., Ltd.).

Herbal materials: Saposhnikoviae Radix (batch No. C25052903, Anguo Runde Pharmaceutical Co., Ltd.); Phellodendri Cortex (batch No. C25062428, Anguo Runde Pharmaceutical Co., Ltd.); Kochiae Fructus (batch No. C24091103, Anguo Runde Pharmaceutical Co., Ltd.); Dictamni Cortex (batch No. 24081004, Anguo Lulutong Traditional Chinese Medicine Decoction Pieces Co., Ltd.); Schizonepetae Herba (batch No. C25020920, Anguo Runde Pharmaceutical Co., Ltd.); Sophorae Flavescentis Radix (batch No. C2511021, Anguo Runde Pharmaceutical Co., Ltd.); Cnidii Fructus (batch No. 250901, Sichuan Wusheng Pharmaceutical Group Co., Ltd.).

Instruments: Thermo Fisher UltiMate 3000 HPLC system equipped with an ultraviolet detector (UVD) and a charged aerosol detector (CAD); UPLC-Triple-TOF/MS system: SCIEX ExionLC™ series high-performance liquid chromatograph (SCIEX, USA), SCIEX X500B QTOF time-of-flight mass spectrometer equipped with an electrospray ionization source (AB SCIEX, USA).

## 3 Experimental

### 3.1 Technical route of this study

In this study, we first compared multiple prompt-engineering strategies to establish the optimal LLM interaction paradigm and screen preliminary chromatographic schemes; subsequently, through progressive experiments of intelligent screening, experimental verification, and stepwise iteration, the final HPLC conditions were determined. On this basis, LC-MS combined with reference-standard spiking experiments was used to qualitatively identify the active constituents of the formula, followed by method validation and quantitative analysis.

The technical route of this study is shown in Figure 1.

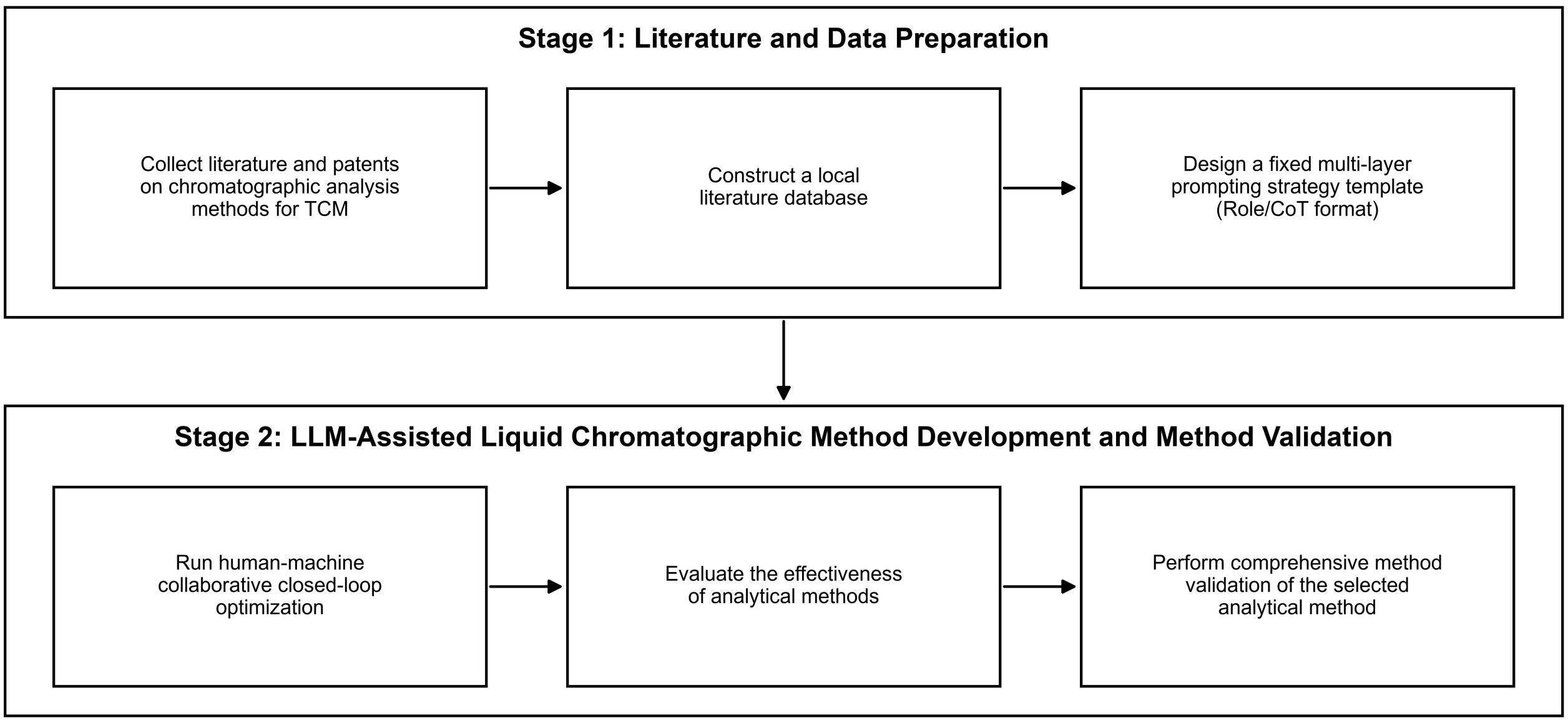

Figure 1 Technical route

### 3.2 Selection of detector

In this study, the ultraviolet absorption characteristics of the target constituents in the seven component herbs of the Eczema External Wash Formula were investigated. Momordin Ic lacks a strong chromophore, so its UV response at 254 nm is significantly weaker than that of the alkaloids, coumarins, and chromones in the formula, making it difficult to ensure quantitative sensitivity and reproducibility by UV detection alone; alkaloids such as dictamnine show peak-shape distortion in UV detection, requiring cross-validation by a non-spectroscopy-dependent detector. Based on these considerations, a combination of UV and CAD detection in series was finally adopted: UV at 254 nm was used to quantify six constituents with good UV response, including berberine hydrochloride[10]; the CAD detector, as a universal detector[11, 12], was dedicated to the quantification of momordin Ic and provided independent chromatographic evidence for peak-purity verification of dictamnine and other constituents.

### 3.3 Preparation of test and reference solutions

#### 3.3.1 Preparation of test solution

The seven herbs (Saposhnikoviae Radix, Phellodendri Cortex, Kochiae Fructus, Dictamni Cortex, Schizonepetae Herba, Sophorae Flavescentis Radix, and Cnidii Fructus) were added to ultrapure water at four times the total mass of the herbs, soaked for 30 min, and decocted for 1 h. The decoction was filtered, and the residue was returned to the decoction pot; ultrapure water at six times the total mass was added and decoction was continued for another 1 h. After filtration, the two filtrates were combined, and the resulting decoction was concentrated at 70 °C using a rotary evaporator to obtain approximately 500 mL of concentrate, which was stored refrigerated for later use.

An aliquot of 1 mL of the Eczema External Wash Formula sample solution was accurately measured and transferred to a 10 mL volumetric flask, methanol was added to volume, and the solution was mixed thoroughly. The mixture was centrifuged at 4600 r/min for 10 min; the supernatant was collected and centrifuged again under the same conditions, and the upper clear liquid was taken as the test solution for later use[13,14].

### 3.3.2 Preparation of reference solution

Appropriate amounts of the reference standards of berberine hydrochloride, dictamnine, magnoflorine, momordin Ic, prim-O-glucosylcimifugin, osthole, and imperatorin were accurately weighed into separate 10 mL volumetric flasks, dissolved in methanol, made up to volume, and mixed to obtain the individual reference standard stock solutions. An accurately measured aliquot of each individual stock solution was then transferred into a single volumetric flask, diluted to the mark with methanol, and mixed to prepare the mixed reference standard solution.

## 3.4 Optimization of HPLC conditions based on large language models

In this study, an artificial intelligence-assisted design strategy was introduced for the development and optimization of HPLC conditions, using the Qwen2.5-14B-V3-pretrain large language model for TCM chromatographic analysis to provide optimization suggestions. The construction method of this large model is described in Supplementary Appendix 6.

### 3.4.1 Comparison of prompt strategies

To systematically evaluate the effectiveness of different prompt-engineering strategies in assisting the generation of HPLC analytical methods for TCMs, five representative prompt schemes were designed and compared in this study (see Supplementary Appendix 1). To objectively quantify the output quality of each scheme, a comprehensive evaluation was conducted from five dimensions: prompt strategy, response quality, professional accuracy, parameter completeness, and core deficiencies[1].

### 3.4.2 Optimal prompt scheme

After comparison, the optimal prompt scheme obtained is as follows:

[Role setting and scenario construction]

You will play the role of a TCM analytical expert with ten years of experience, specializing in using HPLC for fingerprint establishment and multi-component quantification of complex formulas. Your thinking style is rigorous, systematic, focused on deriving experimental schemes from theoretical principles, and you are familiar with the relevant technical requirements of the Chinese

Pharmacopoeia. All your recommendations should be based on chromatographic principles, and you should clearly state their assumptions and limitations.

[Structured information input and task decomposition]

Project background: The TCM formula "Eczema External Wash Formula" is composed of seven herbs: Saposhnikoviae Radix, Phellodendri Cortex, Kochiae Fructus, Dictamni Cortex, Schizonepetae Herba, Sophorae Flavescentis Radix, and Cnidii Fructus, extracted by water decoction.

Key target constituents of interest:

1. Berberine hydrochloride (log P ≈ -1.2, strongly basic quaternary ammonium)

2. Matrine (log P ≈ 1.8, weakly basic)

3. Oxymatrine (log P ≈ -0.4, highly polar)

4. Osthole (log P ≈ 3.8, neutral to weakly polar)

Project objectives:

1. Establish an HPLC method that clearly separates the above four marker constituents and other major unknown peaks;

2. The total run time of the method should be controlled within 60 min;

3. The method must be compatible with conventional C18 columns (e.g., 250×4.6 mm, 5 μm) and a quaternary-pump liquid chromatography system.

[Chain-of-thought and stepwise instructions]

Please think and output strictly in the following order:

Step 1 - Retention behavior assessment: Based on the log P and acid-base properties of the above constituents, analyze their approximate retention strengths on a reversed-phase C18 column, and indicate which constituents may show special chromatographic behavior (e.g., tailing, too weak retention due to high polarity, risk of mutual interference, etc.).

Step 2 - Mobile phase and column recommendation: Based on the above analysis, recommend an initial column type and mobile phase system (e.g., methanol-water or acetonitrile-water; whether a buffer salt is needed, and if so, which buffer salt and pH). Please explain your reasons in detail.

Step 3 - Gradient program and parameter design: Design a complete initial gradient elution program table and explain the logic behind the gradient settings. Also recommend suitable column temperature, flow rate, and detection wavelengths (suggest a full-wavelength scan range and a single quantification wavelength).

Step 4 - Risk prediction: Predict the most likely separation problems under these initial conditions (e.g., high risk of co-elution of two peaks, tailing of a basic constituent, etc.) and propose one or two alternative adjustment strategies in advance.

[Structured output]

Please clearly output your finally recommended initial chromatographic conditions in the following table format, so that they can be directly entered into the instrument control software:

| Parameter | Setting |

| Column | (brand/model, specification) |

| Mobile phase A | (composition) |

| Mobile phase B | (composition) |

| Gradient program | (time (min) & %B) |

| Flow rate | (mL/min) |

| Column temperature | (℃) |

| Detection wavelength | (quantification wavelength and full-scan range) |

| Injection volume | (μL) |

[Method key points]

Meanwhile, please generate a brief "rationale of the method", focusing on explaining why this method can solve the preliminary separation problem of this complex formula. Initial scheme generation and rationale interpretation: combining the role setting of the first layer and the structured input of the second layer, the LLM generates initial chromatographic conditions. The key strategy is to require the model to attach its chain of reasoning. For example: "Acetonitrile-0.1% phosphoric acid aqueous solution is recommended as the mobile phase, because the target constituents matrine (log P = 1.8) and osthole (log P = 3.8) differ greatly in polarity and require gradient elution; adding phosphoric acid can improve the peak shape of basic constituents; the initial 10% acetonitrile ensures the retention of polar constituents."

### 3.5 Experimental verification of chromatographic conditions from the large language model

Considering comprehensively the effects of gradient step, mobile phase acidity, flow rate, and injection volume on separation efficiency, peak shape, and constituent response[15], five initial liquid-chromatographic schemes generated according to the optimal prompt scheme were systematically evaluated from the perspectives of chromatographic kinetics, thermodynamics, and spectral characteristics of the constituents. Finally, three gradient schemes were selected for experimental verification: Scheme 2 (55 min), Scheme 3 (60 min), and Scheme 5 (70 min). Scheme 2 (0.1% formic acid, 1.0 mL/min, 10 μL injection volume) balanced analysis speed and separation efficiency, achieving baseline separation and symmetrical peak shapes for the seven core marker constituents, and showing the best system robustness. Scheme 3 (0.2% glacial acetic acid, 0.5 mL/min, 20 μL injection volume) could significantly improve the signal-to-noise ratio of trace constituents and is suitable for sensitivity optimization; Scheme 5 (0.1% glacial acetic acid, 0.6 mL/min, 20 μL injection volume) provides the largest separation space and is suitable for high-resolution analysis of complex samples, but it shows obvious peak broadening and long analysis time, serving mainly as a reserve scheme for in-depth analysis.

### 3.6 Constituent analysis of the Eczema External Wash Formula based on LC-MS

In this study, LC-MS was used for the qualitative analysis of chemical constituents in the water-extracted concentrate of the Eczema External Wash Formula[16-19]. The liquid-chromatographic conditions employed the chromatographic gradient optimized by the LLM and preliminarily verified as described above. The mass spectrometry used an electrospray ionization source, and full-scan analysis was performed in both positive and negative ion modes to obtain the accurate molecular masses and tandem mass spectrometric fragment information of each constituent in the sample. Qualitative analysis was performed by accurate molecular mass matching and comparison of tandem mass spectrometric fragments with the literature.

### 3.7 Validation and methodological investigation of seven active constituents in the Eczema External Wash Formula

#### 3.7.1 Specificity test

In this study, comprehensive validation was performed from four aspects: exclusion of solvent interference, chromatographic resolution, peak assignment confirmation, and mass spectrometric feature matching. A mixed reference solution and the methanol solvent used for solution preparation were injected for analysis. The chromatographic peak positions of the seven core marker constituents (berberine hydrochloride, dictamnine, magnoflorine, momordin Ic, prim-O-glucosylcimifugin, osthole, and imperatorin) were confirmed by direct comparison of reference-standard retention times and by spiking experiments.

### 3.7.2 Method validation

In this study, the seven core marker constituents were quantified separately using two detectors: berberine hydrochloride, magnoflorine, prim-O-glucosylcimifugin, osthole, and imperatorin have good UV absorption and were quantified using the UV detector at 254 nm; momordin Ic, which lacks a strong chromophore in its molecule and shows weak end absorption in the UV region, was quantified using the CAD detector[11].

3.7.2.1 Linearity

Aliquots of 0.1, 0.15, 0.20, 0.25, 0.30, 0.35, and 0.40 mL of the mixed reference stock solution were accurately measured and placed in 5 mL volumetric flasks, diluted with methanol to volume, and mixed. Linear regression analysis was performed with mass concentration as the abscissa (X) and peak area as the ordinate (Y), and the regression equation and coefficient of determination ($R^2$) were calculated.

3.7.2.2 Precision

The same test solution was injected six times consecutively.

3.7.2.3 Repeatability

Six aliquots of the same batch of test solution were prepared and injected for analysis.

3.7.2.4 Stability

The same test solution was injected for analysis at 0, 2, 4, 8, 12, and 24 h after preparation.

3.7.2.5 Spike recovery test

Nine aliquots of a sample solution of known content were taken and divided into three groups, setting low, medium, and high concentration levels, with the ratio of the amount of reference standard added to the amount of the constituent in the test solution controlled at approximately 0.8:1.0, 1.0:1.0, and 1.2:1.0. For each concentration, three test solutions were prepared in parallel for determination.

## 4 Results and Discussion

### 4.1 Generation of candidate conditions based on the optimal prompt scheme

The generation of the five preliminary liquid-chromatographic schemes and their theoretical rationality evaluation are described in detail in Supplementary Appendix 3.

The chemical constituents of the Eczema External Wash Formula are characterized by a wide polarity range and diverse structural types (e.g., quaternary alkaloids, lipophilic coumarins, and triterpenoid saponins with weak UV absorption). To select optimal chromatographic conditions that balance resolution, sensitivity, and analysis efficiency, five representative initial liquid-chromatographic schemes were designed and preliminarily established with LLM assistance[1,7,20] (Table 1), and their rationality was evaluated. Based on the high-efficiency separation performance of the SunShell C18 (2.6 μm) core-shell column, the initial schemes were constructed for key parameters such as mobile-phase acid modifier, detector type, and gradient logic (Table 1):

Table 1 Initial schemes generated by the LLM

| Comparison dimension | Parameter | Scheme 1 | Scheme 2 | Scheme 3 | Scheme 4 | Scheme 5 |
|---|---|---|---|---|---|---|
| Basic configuration | Mobile phase A | 0.1% formic acid in water | 0.1% formic acid in water | 0.2% glacial acetic acid in water | 0.1% formic acid in water | 0.1% glacial acetic acid in water |
| | Mobile phase B | Acetonitrile | Acetonitrile | Acetonitrile | 0.1% formic acid in acetonitrile | Acetonitrile |
| | Flow rate (mL/min) | 0.60 | 1.00 | 0.50 | 0.80 | 0.60 |
| | Column temperature (°C) | 30.00 | 30.00 | 25.00 | 35.00 | 30.00 |
| | Injection volume (μL) | 10.00 | 10.00 | 20.00 | 5.00 | 20.00 |
| Gradient logic | Total time (min) | 53.00 | 55.00 | 60.00 | 50.00 | 70.00 |
| Detection means | Detector type | CAD | UV (wavelength gradient) | UV (260 nm) | UV (270/325 nm) | UV (280 nm) |
| Design core | | Universal response | Separation-efficiency optimization | Enhanced response of trace constituents | Improved peak shape of basic constituents | Deep separation of isomers |

### 4.2 Experimental verification and iterative optimization of the optimal chromatographic conditions from the large language model

#### 4.2.1 Preliminary chromatographic-behavior evaluation of the three schemes

The chromatograms of the three schemes are shown in Figure S1 of Supplementary Appendix 3. Scheme 2 showed relatively normal chromatographic retention characteristics, with multiple constituents forming independent peaks initially; however, many constituents eluted densely between 10 and 20 min, with local peak clusters and overlap, indicating that the rate of change of the early gradient needed optimization. In Schemes 3 and 5, the constituents were observed experimentally to elute rapidly near the dead time, severely overlapping with the solvent peak. Scheme 2 showed reasonable retention behavior and was determined as the core base scheme for subsequent optimization; Scheme 5, which introduced special parameters such as a long gradient, low flow rate, and large injection volume, still has theoretical value for the detection of trace constituents and was retained as an exploratory scheme, with problem diagnosis and gradient reconstruction performed with the LLM.

### 4.2.2 Fine iteration of Scheme 2

Focusing on the key parameter of the increasing rate of the organic-phase proportion in the gradient program, Scheme 2 was adjusted stepwise three times; the core strategy was to reduce the increasing rate of the organic-phase proportion in the initial stage to widen the elution interval. The three supplementary conditions are shown in Supplementary Appendix 3 (5), and the chromatograms are shown in Figure S3. After three optimizations, all seven constituents were clearly separated within 60 min, with symmetrical peak shapes and a stable baseline. This process demonstrates the advantages of human-AI collaboration: the LLM provides reasonable initial conditions and risk predictions, while the refinement of specific parameters relies on experimental feedback.

### 4.2.3 Determination of the final liquid-chromatographic conditions

The chromatographic conditions were as follows:

Column: SunShell C18 (4.6×150 mm, 2.6 μm); mobile phase: A, 0.1% formic acid aqueous solution; B, acetonitrile; flow rate: 1.0 mL/min; column temperature: 30 °C; detector 1: UV detector at 254 nm; detector 2: CAD detector with nebulizer temperature of 40 °C; data acquisition rate: 10 Hz; filter: 5.0; injection volume: 10 μL; run time: 55 min. The mobile-phase gradient elution conditions are shown in Table 2. The optimal chromatogram is shown in Figure 2.

Table 2 Mobile-phase gradient elution conditions

| Time (min) | Flow rate | A (aqueous, %) | B (organic, %) | Elution curve |
|---|---|---|---|---|
| 0.00 | 1.00 | 95.00 | 5.00 | - |
| 25.00 | 1.00 | 75.00 | 25.00 | Linear |
| 40.00 | 1.00 | 50.00 | 50.00 | Linear |
| 50.00 | 1.00 | 10.00 | 90.00 | Linear |
| 55.00 | 1.00 | 95.00 | 5.00 | Linear (equilibration) |

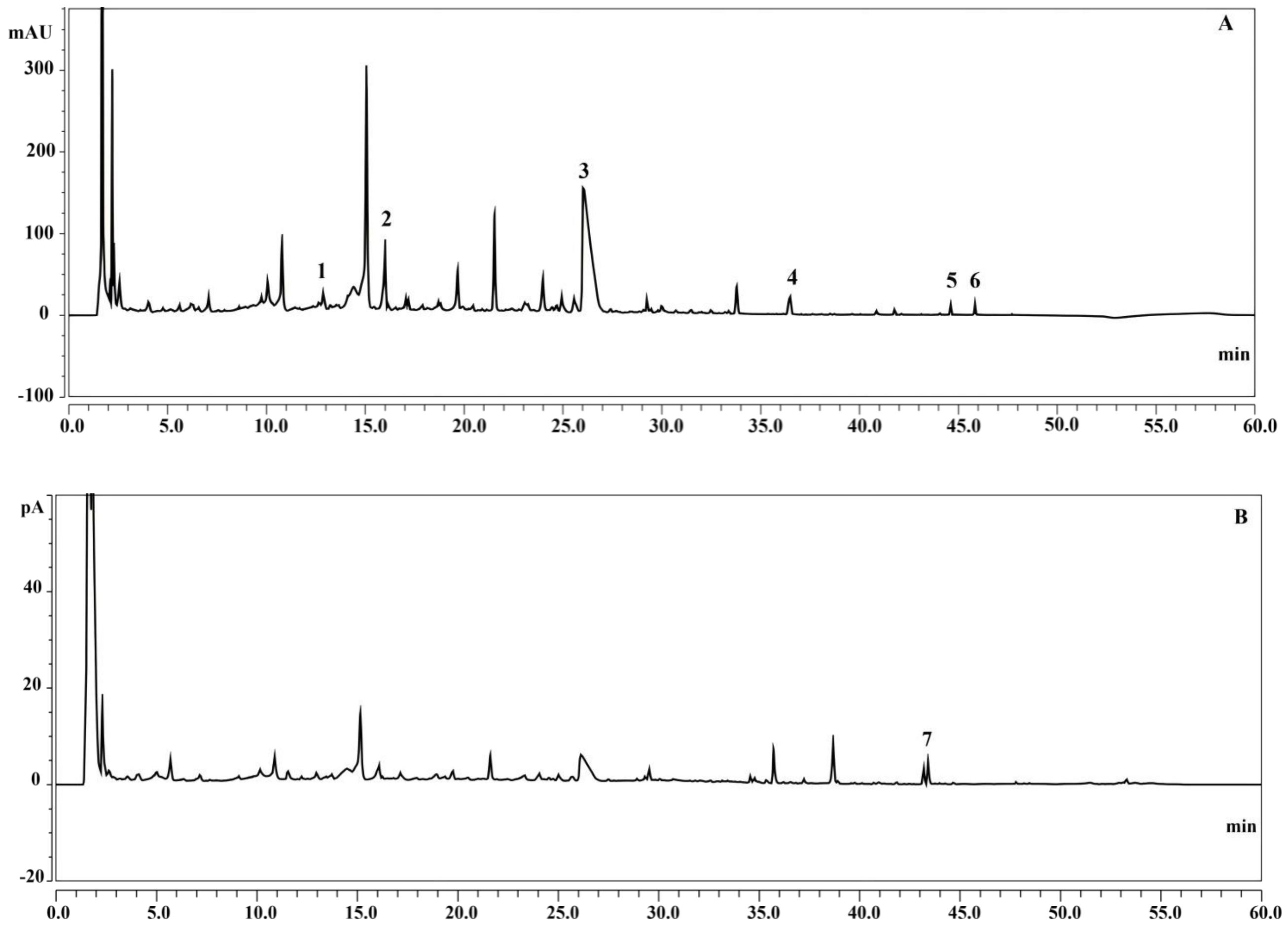

1-magnoflorine, 2-prim-O-glucosylcimifugin, 3-berberine hydrochloride,

4-dictamnine, 5-imperatorin, 6-osthole, 7-momordin Ic

Figure 2 . Optimized HPLC chromatograms obtained with the UV detector (A) and the CAD detector (B)

## 4.3 Constituent analysis of the Eczema External Wash Formula based on LC-MS

### 4.3.1 Mass spectrometric information of the Eczema External Wash Formula

By analyzing the peak shapes of the TIC and comparing the mass-to-charge ratios (m/z) of each constituent with the fragmentation patterns reported in the literature, 20 major chemical constituents were preliminarily identified in the Eczema External Wash Formula; their retention times and mass spectrometric information are listed in Table 3. Detailed analysis is provided in Supplementary Appendix 4.

Table 3 . Summary of spectral interpretation results

| Constituent | Retention time (min) | Ion mode | Measured m/z | Exact Mass (M) | [M+H]+ (m/z) | [M−H]− (m/z) | Error (ppm) | Major fragment ions | Source |
|---|---|---|---|---|---|---|---|---|---|
| Oxymatrine | 1.7 / 3.2 | Pos | 265.1907 | 264.1838 | 265.1911 | 263.1765 | -1.5100 | 247, 205, 148 | Sophorae Flavescentis Radix |
| Oxysophocarpine | 3.0 | Pos | 263.1751 | 262.1681 | 263.1754 | 261.1608 | -1.1400 | 245, 205, 148, 112 | Sophorae Flavescentis Radix |
| Matrine | 1.5 / 2.3 | Pos | 249.1952 | 248.1889 | 249.1962 | 247.1816 | -4.0100 | 148, 150, 98 | Sophorae Flavescentis Radix |
| Sophocarpine | 1.55 / expected >3 | Pos | 247.1798 | 246.1732 | 247.1805 | 245.1659 | -2.8300 | 229, 136, 110, 98 | Sophorae Flavescentis Radix |
| Phellodendrine | 8.5 | Pos | 342.1690 | 342.1700 | 342.1700 | - | -2.9200 | 192, 177, 151 | Phellodendri Cortex |
| Magnoflorine | 10.2 | Pos | 342.1694 | 342.1700 | 342.1700 | - | -1.7500 | 297, 265, 282, 237 | Phellodendri Cortex |
| Prim-O-glucosylcimifugin | 16.0 | Pos | 469.1685 | 468.1632 | 469.1705 | 467.1559 | -4.2600 | 307, 249, 231 | Saposhnikoviae Radix |
| Cimifugin | 19.9 | Pos | 307.1170 | 306.1103 | 307.1176 | 305.1030 | -1.9500 | 249, 189 | Saposhnikoviae Radix |
| Jatrorrhizine | 20.1 | Pos | 338.1383 | 338.1387 | 338.1387 | - | -1.1800 | 323, 280, 265 | Phellodendri Cortex |
| Berberine | 23.0 | Pos | 336.1213 | 336.1231 | 336.1231 | - | -5.3600 | 321, 292, 278, 320 | Phellodendri Cortex |
| Palmatine | 23.6 | Pos | 352.1541 | 352.1543 | 352.1543 | - | -0.5700 | 337, 336, 322, 308 | Phellodendri Cortex |
| Hesperidin | 23.7 | Neg | 609.1826 | 610.1892 | 611.1965 | 609.1819 | 1.1500 | 301, 286, 258 | Schizonepetae Herba |
| Sec-glucosylhamaudol | 26.5 / 29.5 | Pos | 439.1594 | 438.1526 | 439.1599 | 437.1453 | -1.1400 | 277, 259, 241 | Saposhnikoviae Radix |
| Luteolin | 30.5 | Neg | 285.0406 | 286.0477 | 287.0550 | 285.0404 | 0.7000 | 151, 133 | Dictamni Cortex |
| Dictamnine | 36.0 | Pos | 200.0701 | 199.0633 | 200.0706 | 198.0560 | -2.5000 | 185, 157, 130 | Dictamni Cortex |
| Kurarinone | 40 / 46 | Neg | 437.1969 | 438.2042 | 439.2115 | 437.1969 | 0.0000 | 285, 270, 269 | Sophorae Flavescentis Radix |
| Imperatorin | 41 / 43.3 | Pos | 271.0961 | 270.0892 | 271.0965 | 269.0819 | -1.4800 | 203, 175, 153 | Cnidii Fructus |
| Isoimperatorin | 43.3 | Pos | 271.0961 | 270.0892 | 271.0965 | 269.0819 | -1.4800 | 203, 175, 147 | Cnidii Fructus |
| Momordin Ic | 43.6 | Neg | 777.4448 | 778.4498 | 779.4571 | 777.4425 | 2.9600 | 455, 587 | Kochiae Fructus |
| Osthole | 44.5 | Pos | 245.1168 | 244.1099 | 245.1172 | 243.1026 | -1.6300 | 189, 131 | Cnidii Fructus |

### 4.3.2 Determination of quantitative constituents

Based on the TCM formulation theory of monarch, minister, assistant, and guide, this formula aims to clear heat and resolve dampness, dispel wind and relieve itching. Phellodendri Cortex is the monarch herb (clears heat and dries dampness); Sophorae Flavescentis Radix and Dictamni Cortex are the minister herbs (reinforcing dampness-drying and itching relief); Kochiae Fructus, Cnidii Fructus, Schizonepetae Herba, and Saposhnikoviae Radix together are the assistant herbs (dispelling wind and relieving itching, clearing heat and promoting diuresis). Berberine hydrochloride (a highly abundant and active constituent of Phellodendri Cortex), dictamnine and magnoflorine (characteristic constituents of Dictamni Cortex), momordin Ic (a major constituent of Kochiae Fructus), prim-O-glucosylcimifugin (a representative chromone of Saposhnikoviae Radix), and imperatorin and osthole from Cnidii Fructus were selected as the targets for quantitative analysis[21]; their chemical classes and biological activities are shown in Table 4.

Table 4 Selection of the seven core constituents

| Herb | Formulation role | Marker constituent | Chemical class | Biological activity |
| --- | --- | --- | --- | --- |
| Phellodendri Cortex | Monarch | Berberine hydrochloride | Benzylisoquinoline alkaloid | Has significant antibacterial, anti-inflammatory, heat-clearing and dampness-drying effects, and is the material basis of the core heat-clearing effect of the formula[22]. |
| Dictamni Cortex | Minister | Dictamnine | Quinoline alkaloid | Has antifungal, anti-inflammatory, and antipruritic effects, and is a representative active component for the clinical treatment of skin eczema and urticaria[23,24]. |
| Phellodendri Cortex / Dictamni Cortex | Minister | Magnoflorine | Aporphine alkaloid | Has immunomodulatory, anti-inflammatory, and antioxidant effects, and is commonly used as a quality-control marker for polar alkaloids in Dictamni Cortex[25,26]. |
| Kochiae Fructus | Assistant | Momordin Ic | Triterpenoid saponin | Has significant heat-clearing, diuretic, anti-allergic, and antipruritic effects; combined use with Dictamni Cortex can enhance the wind-dispelling and itching-relieving effect |
| Saposhnikoviae Radix | Assistant | Prim-O-glucosylcimifugin | Chromone | Exerts wind-dispelling, exterior-releasing, dampness-overcoming, and pain-relieving effects, can synergistically regulate the skin surface environment, and assists the main herbs in eliminating exogenous pathogenic factors[29]. |
| Cnidii Fructus | Assistant | Osthole | Simple coumarin | Has strong antipruritic, insecticidal, and anti-allergic effects, and is one of the core pharmacodynamic constituents of the formula targeting eczema itching symptoms[30,31]. |
| Cnidii Fructus | Assistant | Imperatorin | Linear furanocoumarin | Has analgesic, anti-inflammatory, and anti-allergic effects, and together with osthole forms the coumarin group used to evaluate the quality of Cnidii Fructus[32,33]. |

### 4.4 Validation and methodological investigation of the seven constituents of the Eczema External Wash Formula

#### 4.4.1 Linearity results

The regression equations and linear ranges of berberine hydrochloride, magnoflorine, prim-O-glucosylcimifugin, osthole, imperatorin, and momordin Ic are shown in Table 5. All seven constituents showed the following linear relationships within their respective concentration ranges, with coefficients of determination greater than 0.99.

Table 5 Linearity of each constituent

| Constituent | Regression equation | $R^2$ | Linear range (μg/mL) | Detector type |
|---|---|---|---|---|
| Berberine hydrochloride | y = 0.7018x | $R^2 = 0.9981$ | 19.63~137.41 | UV |
| Dictamnine | y = 0.2917x | $R^2 = 0.9944$ | 4.66~37.26 | UV |
| Prim-O-glucosylcimifugin | y = 0.4159x | $R^2 = 0.9904$ | 6.82~54.53 | UV |
| Momordin Ic | y = 0.0691x | $R^2 = 0.9911$ | 1.64~11.48 | CAD |
| Magnoflorine | y = 0.1036x | $R^2 = 0.9978$ | 10.80~86.39 | UV |
| Imperatorin | y = 1.1919x | $R^2 = 0.9952$ | 0.21~23.4 | UV |
| Osthole | y = 0.2425x | $R^2 = 0.9931$ | 0.31~26 | UV |

#### 4.4.2 Precision, repeatability, and stability results

The same test solution was injected six consecutive times for the precision test, and the RSD of the content of each marker constituent was less than 2.0%. Six test solutions were prepared in parallel for the repeatability test, and the RSD of the content of each constituent was less than 3.0%. The test solution was left at room temperature for 0–24 h for the stability test, and the RSD of the content of each constituent was ≤4.0%. These results indicate good instrument precision and method repeatability, and the test solution is essentially stable within 24 h.

Table 6 Results of precision, stability, and repeatability tests for each constituent

| Constituent | Precision RSD (%) | Stability RSD (%) | Repeatability RSD (%) |
|---|---|---|---|
| Berberine hydrochloride | 1.85 | 2.36 | 2.52 |
| Dictamnine | 1.90 | 2.88 | 2.67 |
| Osthole | 1.74 | 2.50 | 1.86 |
| Prim-O-glucosylcimifugin | 1.97 | 3.18 | 1.99 |
| Momordin Ic | 1.73 | 2.72 | 2.24 |
| Magnoflorine | 1.63 | 2.23 | 2.83 |
| Imperatorin | 1.57 | 3.14 | 2.42 |

### 4.4.3 Spike recovery results

Three spiking levels of 80%, 100%, and 120% were set, and the spike recovery of each constituent was calculated; the results are shown in Table 7. The average recoveries of most marker constituents ranged from 92.95% to 108.53%, indicating good accuracy of the method.

Table 7 Results of the spike recovery test

| Constituent | Low level (80%) | | Medium level (100%) | | High level (120%) | |
|---|---|---|---|---|---|---|
| | Average recovery (%) | RSD(%) | Average recovery (%) | RSD(%) | Average recovery (%) | RSD(%) |
| Magnoflorine | 98.08 | 1.19 | 98.58 | 1.88 | 102.29 | 2.37 |
| Prim-O-glucosyl | 96.31 | 1.37 | 94.61 | 0.78 | 102.82 | 2.51 |
| Berberine | 100.98 | 1.73 | 100.57 | 1.70 | 101.13 | 0.69 |
| Dictamnine | 97.29 | 1.96 | 98.24 | 2.20 | 97.73 | 0.78 |
| Momordin Ic | 96.32 | 0.66 | 95.99 | 1.69 | 97.62 | 1.54 |
| Imperatorin | 94.45 | 0.71 | 95.29 | 2.58 | 92.95 | 3.03 |
| Osthole | 108.20 | 2.74 | 108.53 | 1.40 | 107.61 | 1.28 |

## 4.5 Application of the method

The contents of the marker constituents in the water extracts of the Eczema External Wash Formula from 10 batches were determined by the established HPLC method. The preparation of the water extract and the test solution was the same as that described in Section 3.3.1, and the results are shown in Table 8. The mass concentrations of magnoflorine, prim-O-glucosylcimifugin, berberine hydrochloride, dictamnine, momordin Ic, imperatorin, and osthole in the water extracts were 57.25–64.90, 33.68–49.91, 174.38–255.99, 24.02–27.49, 20.14–22.03, 3.53~4.47 and 17.16–20.52 μg/mL, respectively. Overall, the contents of the marker constituents varied among batches, and the mass concentration of berberine hydrochloride was relatively higher than those of the other marker constituents.

Table 8 Mass concentrations of the seven marker constituents in water extracts of the Eczema External Wash Formula from different batches (μg/mL)

| Batch No. | Magnoflorine | Prim-O-glucosyl cimifugin | Berberine hydrochloride | Dictamnine | Momordin Ic | Imperatorin | Osthole |
|---|---|---|---|---|---|---|---|
| S1 | 62.53 | 49.91 | 255.99 | 27.49 | 24.50 | 4.47 | 20.52 |
| S2 | 62.63 | 36.93 | 200.81 | 27.21 | 23.23 | 4.28 | 20.17 |
| S3 | 57.25 | 42.43 | 221.15 | 27.10 | 24.91 | 3.84 | 18.58 |
| S4 | 64.90 | 36.28 | 179.36 | 25.70 | 21.13 | 4.12 | 18.57 |
| S5 | 58.96 | 33.68 | 176.79 | 24.57 | 21.86 | 3.66 | 18.08 |
| S6 | 64.75 | 51.81 | 253.38 | 28.33 | 25.31 | 4.25 | 18.92 |
| S7 | 62.99 | 37.02 | 200.02 | 26.96 | 25.54 | 4.10 | 19.21 |
| S8 | 61.11 | 44.80 | 225.64 | 26.95 | 22.58 | 3.71 | 17.91 |
| S9 | 63.81 | 36.56 | 178.37 | 24.91 | 21.75 | 3.91 | 17.73 |
| S10 | 58.50 | 33.86 | 174.38 | 24.02 | 20.14 | 3.53 | 17.16 |

### 4.6 Discussion

In this study, using the Eczema External Wash Formula as the target analyte, we attempted to establish a human–machine collaborative HPLC method-development model of "LLM recommendation—experimental verification—iterative optimization". The comparison of five prompt strategies showed that the direct-instruction strategy misidentified berberine hydrochloride as a flavonoid and recommended an incorrect detection wavelength. The open-interaction strategy exhibited basic conceptual errors such as using ion-exchange chromatography to separate neutral coumarins, while the log P-only strategy lacked core method parameters. The "expert role + structured chain-of-thought" strategy accurately identified the silanol adsorption risk of basic constituents and recommended a reasonable mobile phase system, indicating that structured logical guidance plays a decisive role in suppressing LLM knowledge hallucinations. The human–machine collaborative model in this study both avoids the uncontrollability of relying entirely on LLMs and overcomes the reliance on experience in purely manual development.

This model still has several limitations: (1) There is a gap between LLM predictions and actual chromatographic behavior. (2) Parameter fine-tuning mainly relies on human experience. (3) This model has so far only been validated on a single formula. (4) In this study, UV detection used a single wavelength of 254 nm, but the maximum absorption of coumarin constituents may be above 300 nm. Therefore, switching to DAD multi-wavelength detection may further optimize sensitivity. (5) This study used only a single LLM. If more LLMs were compared, an LLM more suitable for assisting the establishment of HPLC methods might be selected.

**5 Conclusions**

This study verified the feasibility of LLM-assisted HPLC method development. By comparing five prompt-engineering strategies, it was confirmed that the "expert role + structured chain-of-thought (CoT)" strategy can generate theoretically reasonable initial chromatographic conditions, effectively narrowing the screening range of method development. After stepwise iterative optimization, the final chromatographic conditions were determined: a SunShell C18 column (4.6×150 mm, 2.6 μm); gradient elution with a 0.1% formic acid aqueous solution–acetonitrile system at a flow rate of 1.0 mL/min and a column temperature of 30 °C; seven target compounds could be effectively separated within 55 min. Twenty chemical constituents were identified by LC-MS, and based on the "monarch, minister, assistant, and guide" principle, berberine hydrochloride, dictamnine, magnoflorine, momordin Ic, prim-O-glucosylcimifugin, osthole, and imperatorin were selected as the marker constituents. In method validation, the linearity, stability, precision, repeatability, specificity, and spike recovery were all satisfactory. The LLM-assisted method-development model proposed in this study is expected to be extended to the screening of HPLC conditions for other TCM formulas.

**Funding:**

This research was supported by the Science and Technology Co-construction Project of the National Comprehensive Reform Demonstration Zone for Traditional Chinese Medicine (Project No. GZY-KJS-ZJ-2025-053).

**Institutional Review Board Statement:** Not applicable

**Informed Consent Statement:** Not applicable

**Data Availability Statement:** The data that support the findings of this study are available from the corresponding author upon reasonable request.

**Conflicts of Interest:** The authors declare no conflict of interest